\documentclass{PoS}

\usepackage{graphicx}
\usepackage{color}

\newcommand{\ksea}{\mbox{$\kappa^{\rm sea}$}}

\newcommand{\be}{\begin{equation}}
\newcommand{\ee}{\end{equation}}

\newcommand{\bea}{\begin{eqnarray}}
\newcommand{\eea}{\end{eqnarray}}
\newcommand{\NN}{\\ \nonumber}

\def\gtap{\;\raisebox{-.5ex}{\rlap{$\sim$}} \raisebox{.5ex}{$>$}\;}

\newcommand{\er}[2]{{\scriptstyle {}^{+{#1}}_{-{#2}}}}

\newcommand{\insertFig}[4]{
\begin{figure}
\begin{center}
\includegraphics[width=#4\textwidth]{#1}
\caption{#2}
\label{#3}
\end{center}
\end{figure}
}

\PoS{PoS(LAT2005)049}

\title{Chiral and Continuum Extrapolation of Partially-Quenched Hadron Masses}

\ShortTitle{Extrapolations of Partially-Quenched Masses}

\author{\speaker{C. R. Allton}, W. Armour, \\
Department of Physics, University of Wales Swansea, Swansea SA2 8PP, UK}
\author{D. B. Leinweber,\\
CSSM and Department of Physics, University of Adelaide, Adelaide 5005, Australia}
\author{A. W. Thomas, R. D. Young\\
Jefferson Laboratory, Newport News, VA 23606, USA}

\abstract{
Using the finite-range regularisation (FRR) of chiral effective field
theory, the chiral extrapolation formula for the vector meson mass is
derived for the case of partially-quenched QCD.  We re-analyse the
dynamical fermion QCD data for the vector meson mass from
the CP-PACS collaboration.
A global fit, including finite lattice spacing effects,
of all 16 of their ensembles is performed.
We study the FRR method together with a naive polynomial approach
and find excellent agreement ($\sim 1$\%) with
the experimental value of $M_\rho$ from the former approach.
These results are extended to the case of the nucleon mass.
}

\FullConference{XXIII International Symposium on Lattice Field Theory\\
25-30th July 2005\\
School of Mathematics, Trinity College Dublin}

\begin{document}


\section{Introduction}

Current computers and algorithms enable lattice gauge theorists to
perform simulations of QCD over a relatively wide range of parameters,
even in the dynamical-fermion case. However, despite the enormous progress
made, physical values of the light quark masses are still not within
reach. In order for the lattice to make predictions at these quark
masses, a chiral extrapolation procedure is inevitable.

The traditional (dimensional regularised) chiral perturbation theory approach
is well known to suffer from convergence problems. As a symptom of this,
very poor agreement is found when lattice data is fitted to
a truncated (dim-reg) chiral series using the known values of the
non-analytic (in quark mass) coefficients \cite{derek}.
This situation can be understood on physical grounds due to the fact that
chiral perturbation theory is an effective theory. When effective theories
are applied beyond their range of validity, more and more interaction
terms are typically required to mop up for the physics that the effective
theory intrinsically misses.
Such a situation occurs in the case of chiral perturbation theory as the quark
mass becomes moderate.
In this case, as the light quantum fields (i.e. the pions)
become massive and their Compton wavelengths decrease in size, they begin to
probe the quark nature underlying the hadronic fields in the effective theory.
The chiral effective theory therefore breaks down.

Fortunately we have comprehensive lattice data over a wide range of
quark masses in the region where the chiral perturbation theory is no longer
valid. This data tells us that the hadonic masses are typically
very linear in the quark mass, $m_q$, for $m_q \gtap m_s/2 \approx 50$
MeV.

The aim of using chiral extrapolation techniques is therefore to
respect both the properties of the effective theory at small $m_q$,
and the lattice data at moderate to large $m_q$.  The finite-range
regulator (FRR) or Adelaide approach to chiral effective theory
provides such a method.  In this approach, a scale, $\Lambda$, is
introduced representing the transition scale where pion
interaction effects should be moderated.  This scale is introduced naturally
in a form factor for the pion-hadron interactions.
In this formalism, pion loops are moderated like $\Lambda\!/\!m_q$
as $m_q$ grows, {\em and} FRR reproduces (by construction) the
structure of traditional dim-reg chiral effective theory
in the $m_q \rightarrow 0$ limit. The FRR approach
therefore provides a marriage between chiral effective theory and
lattice data.

In this talk, we extend previous FRR work to the vector meson mass
in the ``partially-quenched'' case (i.e. when $m_q^{sea} \ne m_q^{val}$).
This method is then applied to a dataset from the CP-PACS
Collaboration \cite{cppacs}.
An extension of this work to the partially-quenched nucleon mass
is also briefly discussed.
A full description of this work can be found in \cite{us_letter, us_paper}


\section{The FRR Approach}

We introduce the nomenclature $M_{V(PS)}(\beta,s;v_1,v_2)$ for the vector (pseudoscalar)
meson mass calculated with a gauge coupling of $\beta$, sea quark mass of $s$,
and valence quark masses of $v_1$ and $v_2$.

The FRR expression for the vector mass is \cite{us_letter, us_paper}
\be
M_V^{2}(\beta,s;v,v) =
(a_0 + a_2 M^2_{PS}(\beta,s;v,v) + a_4 M^4_{PS}(\beta,s;v,v) + \ldots)^2
+ \Sigma_{TOT}
\label{eq:adelaide}
\ee
\bea \nonumber
\mbox{where }\;\;\;\;\;\;
\Sigma_{TOT} &=&
    \Sigma^\rho_{\pi\pi}   (M^2_{PS}(\beta,s;s,v))
 +  \Sigma^\rho_{\pi\omega}(M^2_{PS}(\beta,s;s,v)) \NN
&+& \Sigma_{DHP}(M^2_{PS}(\beta,s;s,v), M^2_{PS}(\beta,s;v,v),M^2_{PS}(\beta,s;s,s)).
\eea
Note that we derive this expression for vector mesons composed only of
{\em degenerate} valence quarks.

The $\Sigma$ terms are self-energies and are diagrammatically depicted in
Fig.~1 of \cite{us_letter}.
We give below the definition of $\Sigma^\rho_{\pi\omega}$; see \cite{us_letter} for
expressions for $\Sigma^\rho_{\pi\omega}$ and $\Sigma_{DHP}$.
\bea \nonumber
{\Sigma^{\rho}_{\pi\omega}} &=&
-\frac{g_{\omega\rho\pi}^{2}\mu_{\rho}}{12\pi^{2}}\int_{0}^{\infty}
\frac{k^{4}u_{\pi\omega}^{2}(k)~dk}{\omega_{\pi}^{2}(k)} 
\NN
\mbox{with }\;\;\;\;\;\;\;
\omega_{\pi}^{2}(k) &=& k^{2} + M^2_{PS}(\beta,s;s,v)
\;\;\;\;\;\mbox{and }\;\;\;\;\;\;\;
u(k) = \frac{\Lambda^{4}}{(\Lambda^{2}+k^{2})^2}.
\eea
Here $g_{\omega\rho\pi} \mu_\rho$ defines the coupling of $\rho \to
\omega \pi$ with $g_{\omega\rho\pi} = 16$ GeV$^{-1}$
and the physical $\rho-$mass $\mu_\rho=770$ MeV.
The form factor $u(k)$ has a natural scale, $\Lambda$, which was
referred to in the previous section.
Finally note that Eq.~(\ref{eq:adelaide}) reproduces dim-reg
$\chi$PT as $M_\pi \rightarrow 0$.

The momentum integrals in the $\Sigma$ definitions are discretised to
the lattice using
\[
4 \pi \int_{0}^{\infty}k^2dk =
\int d^3k \approx \frac{1}{V}\left(\frac{2 \pi}{a}\right)^3 \sum_{k_x,k_y,k_z}
\]
with $k_\mu$ being constrained by the finite periodic volume of the lattice.


\section{CP-PACS Data}
\label{sec:cppacs}

Data for the vector meson mass was taken from the CP-PACS
Collaboration \cite{cppacs}.
These simulations used two-flavours of mean-field improved Wilson fermions
with Iwasaki glue.
There is, therefore, the possibility of some residual ${\cal O}(a)$
systematics.
The CP-PACS dataset comprises of four different $\beta$'s,
with four $\ksea$ values for each $\beta$ value,
making a total of 16 independent ensembles.
In addition, each ensemble has five different valence quarks, meaning
that there are a total of 80 (degenerate) meson masses.
We represent each of these masses by a 1000-element bootstrap
ensemble which is Gaussianly-distributed about the CP-PACS
central value with the FWHM set to the quoted error in \cite{cppacs}.
The parameter space spanned by this CP-PACS dataset is displayed in
Fig.~\ref{fig:cppacs}.
\insertFig{fg_a_vs_mps2_new.eps}
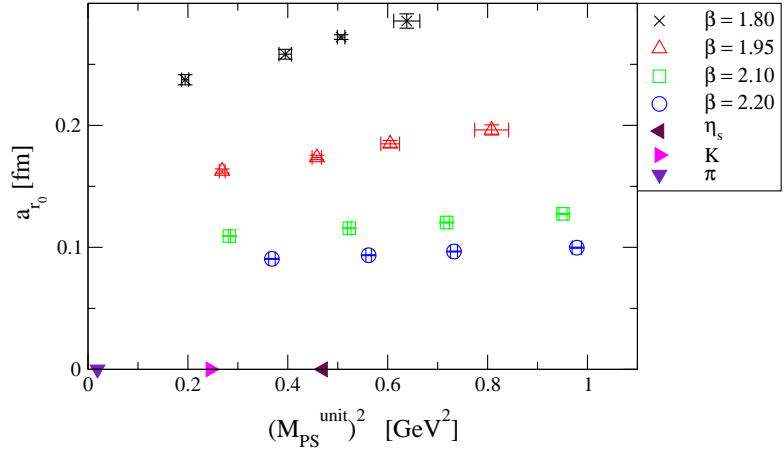
{CP-PACS parameter range in lattice spacing, $a$ (from $r_0$) and
$M_{PS}^{unit}$, where ``unit'' refers to the unitary point
(i.e. $m_q^{sea}=m_q^{val}$). The physical
light pseudoscalar mesons are included for reference.}
{fig:cppacs}
{0.68}


\section{Fitting Procedure}
\label{sec:fit}

The procedure used to fit the CP-PACS data to the FRR functions is
as follows.
Prior to the extrapolation, we first convert all masses into physical units
using both the string tension, and hadronic scale, $r_0$.
This has the following two advantages compared to an analysis
with dimensionless hadron masses:
different ensemble's data can be combined together in a global fit; and
since dimension{\em ful} mass predictions from lattice simulations
are {\em ratios} of lattice mass estimates, some systematic and
statistical errors will cancel.

We then use the FRR functional form below to fit the data.
\be
\sqrt{M_V^{2}(\beta,s;v,v)
\!-\! \Sigma_{TOT}}
\!=\! (a^{cont}_0\!\!\! +\! X_1 a \!+\! X_2 a^2) \!+\! a_2 M^2_{PS}(\beta,s;\!v,v)
 \!+\! a_4 M^4_{PS}(\beta,s;\!v,v) \!+\! a_6 M^6_{PS}(\beta,s;\!v,v).
\label{eq:global}
\ee
This is simply Eq.~(\ref{eq:adelaide}) modified by the addition of
${\cal O}(a,a^2)$ terms to the $a_0$ coefficient
in order to model the lattice spacing systematics.
We tested the addition of terms ${\cal O}(a,a^2)$ to the $a_{2,4,6}$
coefficients as well,
but found that these terms were either compatible with zero,
or lead to unstable fits.
Variants of Eq.~(\ref{eq:global}) were used to study the
stability of the fit procedure:
including the $X_1 a$ term, or alternatively setting $X_1\equiv 0$;
and including the ``cubic'' $a_6 M_{PS}^6$ term, or alternatively
setting $a_6\equiv 0$ (the ``quadratic'' fit).

As well as the FRR function in Eq.~(\ref{eq:global}), we fitted the data
to a naive polynomial fit function defined by Eq.~(\ref{eq:global})
but with $\Sigma_{TOT}$ set to zero (labelled as {\em Naive} in the following).

The above fitting functions were used in a ``global'' fit of all the CP-PACS
data. Since there are 80 meson mass datapoints (see Sec.~\ref{sec:cppacs})
a highly constrained fit was obtained since the number of fitting
parameters is tiny compared to 80. 
See \cite{us_letter, us_paper} for details of the fit procedure

Recall that in the case of the FRR fit, an extra parameter, $\Lambda$ is
introduced to govern the sum of higher-order terms of the expansion with
guidance from lattice QCD data.
We find that $\Lambda$ is constrained well by the large CP-PACS data set.
Figure~\ref{fig:lambda} (upper plot) shows the behaviour of the $\chi^2$
for the above FRR fits.
Results for the four variants to Eq.~(\ref{eq:global}) as discussed above
are shown.
The insert shows the result for the preferred fitting function choice
(which is $X_1 \equiv 0$ and $a_6 \equiv 0$, see \cite{us_paper}).
By increasing $\chi^2$ by one from its minimum value, we obtain the
acceptable range of $\Lambda = 655(35)$ MeV.

\insertFig{fg_vs_lambda.eps}
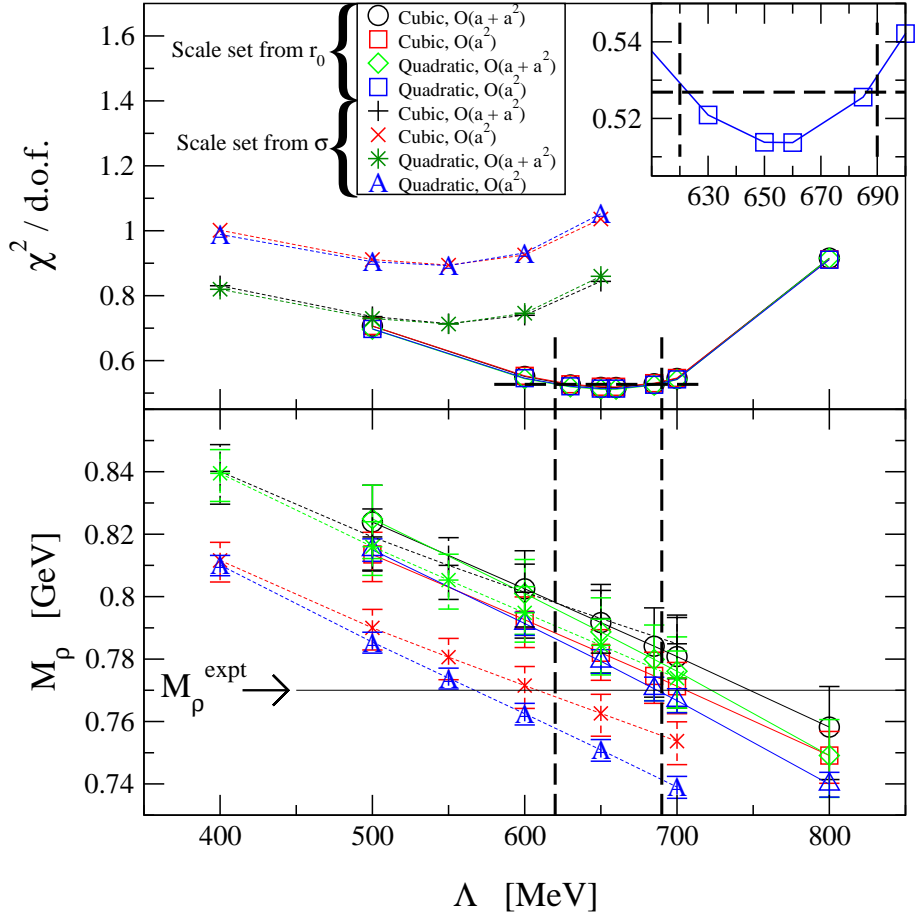
{Upper: $\chi^2/d.o.f.$ versus the FRR $\Lambda$ parameter.
(The insert shows a close up of the $\chi^2$ minimum region.)
Lower: $M_\rho$ prediction versus $\Lambda$.
The vertical dashed lines represent $\Lambda=655(35)$ MeV.}
{fig:lambda}
{0.8}

We find that the FRR approach gives lower $\chi^2$ values than
the naive polynomial fits; that the chiral series (in the FRR
approach) is very convergent (i.e. the inclusion of the ${\cal
O}(M_{PS}^6)$ term is irrelevant in these fits);
and that $r_0$ is a more stable quantity to set the scale than
the string tension.

The success of the FRR approach in fitting the data can be seen
in Fig.~\ref{fig:mv}
by comparing the raw CP-PACS data in the upper plot with 
the data corrected according to Eq.~(\ref{eq:global}) in the lower plot.

\insertFig{fg_raw_famous.eps}
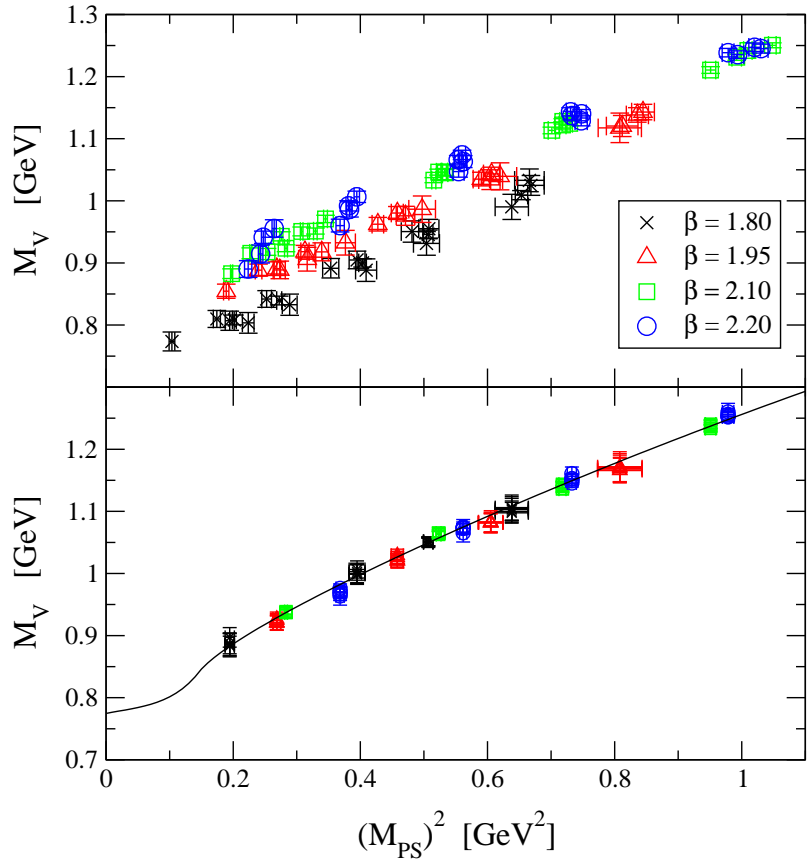
{Upper: The raw 80 CP-PACS data.
Lower: The same data after correction according to Eq.~(4.1).}   
{fig:mv}
{0.7}


\section{$M_\rho$ and $M_n$ estimate}

The final value estimate for $M_\rho$ is obtained using the
$a_{0,2,4}$ parameters from the fit, and by setting
$a=0$, and $M_{PS}$ to the physical pion mass.
We obtain
\be
M_\rho^{FRR} = 778(4)\er{13}{0}\er{8}{9} \;\;\textrm{MeV}
\;\;\;\;\;\mbox{and}\;\;\;\;\;
M_\rho^{Naive} = 825(4)\er{12}{8} \;\;\textrm{MeV}
\ee
where the errors are statistical, fit procedure, and, in the
FRR case, from the variation in $\Lambda$.
This last error is obtained from the lower plot in Fig.~\ref{fig:lambda}, where the
variation of the $M_\rho$ prediction as a function of $\Lambda$
is shown. Using the range $\Lambda = 655(35)$ MeV (see Sec.~\ref{sec:fit})
we obtain the third error. 

Note that we do not include an error estimate due to
the uncertainty in $r_0$ itself, nor from
the fact that the number of dynamical flavours, $N_f \ne 2+1$.

As can be seen the errors from the various sources are typically around 1\%,
including the uncertainty due to the $\Lambda$ value.
Note also that the FRR $M_\rho$ estimate is in comfortable agreement
with the experimental value (in contrast with the naive polynomial
fit).

In \cite{cppacs}, nucleon mass data was also published. The FRR formula
for the partially-quenched nucleon case was derived and a similar
fitting procedure to that outlined here for the vector case was applied.
The estimate of the nucleon mass from this method is
(full details will appear in \cite{us_paper})
\[
M_{N}^{FRR} = 965(15)\er{41}{0}\er{13}{8} \;\;\textrm{MeV}
\;\;\;\;\;\mbox{and}\;\;\;\;\;
M_{N}^{Naive} = 1023(15)\er{53}{0} \;\;\textrm{MeV}.
\]
In this case the agreement with the physical value is reasonable in the
FRR case (and non-existent in the naive polynomial case).
Again, the dependency of $M_N$ on $\Lambda$ is modest.


\section{Conclusions}

We have generalised the FRR chiral approach to ``partially-quenched''
vector meson and nucleon mass cases, obtaining continuum estimates of
$M_\rho$ and $M_N$ which are in agreement with experimental values.
The estimated systematic errors in the masses due to the fit procedure
are $\sim 1-2$\%.
The FRR approach was found to be ``model independent'' in the sense
that the physical predictions' dependencies on the $\Lambda$-parameter
was found to be of the same order as the other systematic error sources.





\end{document}